\documentclass[journal]{IEEEtran}
\usepackage[T1]{fontenc}
\usepackage{lmodern}
\usepackage{stackengine}

\usepackage{amsmath}

\usepackage{stackengine}

\usepackage{wrapfig}
\usepackage{caption}
\usepackage[framemethod=tikz]{mdframed}
\usetikzlibrary{calc}
\usepackage{kantlipsum}

\title{Modelling Quantum Channels Carrying Classical Information}
\author{Indrakshi Dey, National University of Ireland, Maynooth, indrakshi.dey@mu.ie\\
Simon L. Cotton, Queen’s University Belfast, UK, simon.cotton@qub.ac.uk}
\date{December 2021}

\begin{document}

\maketitle

\begin{abstract}
 We use the concept of coupled quantum harmonic oscillators to model the propagation environment in which a quantum link carrying either classical or quantum information operates. Using the analogy between the paraxial optical wave equation and the stationary Schrödinger equation and applying the Caldirola-Kanai Hamiltonian for solving the time-dependent Schrödinger equation; we calculate the propagation field strength and the corresponding average received signal energy.
\end{abstract}

\section{Introduction}

Photons are natural carriers of information in fiber and free-space optical networks. Since photons mediate the fundamental forces of nature, understanding information flows in photonic systems is of both theoretical and practical importance. Noise in conventional networks is often well-approximated by additive Gaussian noise, and the seven decades since Shannon’s theory was introduced have seen the emergence of a mature theory of communication in such practical networks. Low power optical noise, on the other hand, is quantum mechanical in nature. Despite the considerable progress made recently, comparatively simple questions governing the point-to-point capacities of finite-dimensional and arbitrary-dimensional realistic quantum channels remain largely unanswered.

The entropic measure of information content within a transmitted message/data is termed as mutual information. The maximum amount of mutual information that can flow per second from transmitter to receiver is defined as channel capacity, $C$. The capacity of practical communication channels depends on i) noise introduced by transmitter, receiver, and the propagation medium and ii) channel properties that affect the transmitted information differently over large and small timescales. 

As human beings, we generate classical information, and we can only understand information in classical format. To realize the full potential of quantum communication systems when used to transfer classical information, we need to encode classical information into a set of quantum states at the input of quantum channel and decode those quantum states back to classical information through measurement at the output of the quantum channel. To answer the generic question of what the capacity of such a channel will be, let us assume a single linear quantum channel. To calculate $C$ for a single linear quantum channel, we first need to relate the communication theorist’s concept of a channel to the physicist’s description of a quantum field. Considering this, a channel can be defined as the medium or vacuum through which an electromagnetic field propagates. The propagating field is referred to as the signal. An electromagnetic field encompasses electromagnetic waves that oscillate longitudinally in space pushing energy to flow in a transverse direction. To map the electromagnetic field to the quantum field, we need to differentiate between the transverse and longitudinal modes of a quantum field \cite{1}. 

The transverse mode can be characterized by its position in space in a direction perpendicular to the propagation of energy and spin state or polarization. The longitudinal mode is characterized by its position in space along the direction of propagation and its position in time. Therefore, a transverse mode can contain multiple longitudinal modes \cite{2}. Now, both for electromagnetic and quantum fields, it is possible to separate the orthogonal transverse mode unambiguously at the channel output; a feature always used in classical communication systems for detecting information at the receiver. To that end, it is possible to transmit information independently over each transverse mode at the transmitter side and therefore, each transverse mode of a quantum field can be assumed as an independent communication channel. So, when dealing with the capacity of a single channel, we do not need to think about the transverse properties at all. We just need to distinguish between the longitudinal modes arriving at the receiver and $C$ can be calculated as the maximum number of distinguishable longitudinal modes that can arrive at the output of the channel over a certain duration of time and bandwidth of operation. A simple diagrammatic representation of this correspondence between quantum and classical channels is provided below.
\begin{figure*}[t]
\begin{center}
\includegraphics[width=0.8\linewidth]{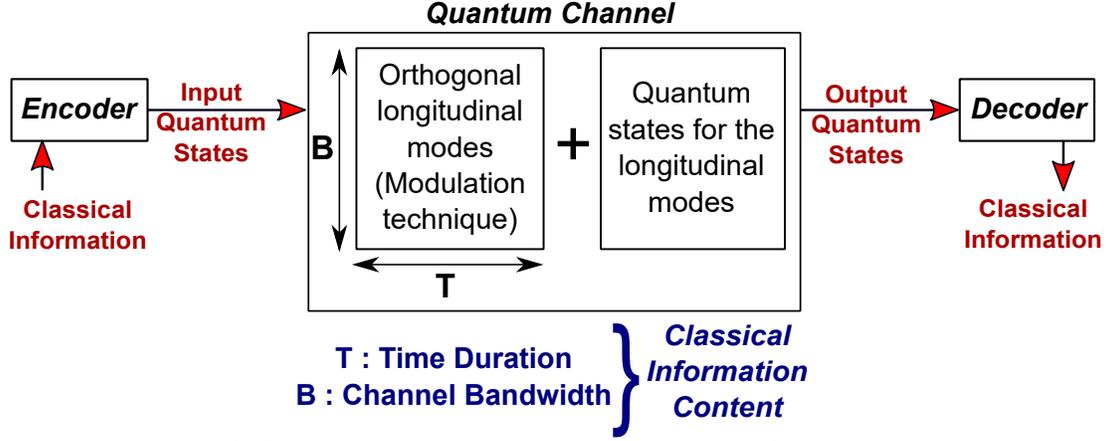}
\end{center}
\vspace*{-4mm}
\caption{Correspondence between quantum field and classical channel.}
\label{fig3}
\vspace*{-2mm}
\end{figure*}

\section{Background}

The capacity of a classical channel carrying classical information is given by Shannon’s theorem \cite{3}, 
\begin{equation}
 C = B\log_{2}\Big(1+\frac{\Upsilon}{N_{0}B}\Big)  \label{eq1}
\end{equation}
where $\Upsilon$ is the transmit signal power for an input that is time-limited and band-limited to $B$Hz and $N_0$ is the power spectral density of the channel noise that is Gaussian distributed. The input and output waveforms of the channel are also assumed to be Gaussian distributed. Classical information can be transmitted over a bosonic quantum channel by encoding the information onto the Fock states (the state of $`n'$ number of photons that can be used to represent a light pulse). If Fock states are used, the maximum rate at which a quantum channel can transmit classical information without any error is given by \cite{4}, 
\begin{equation}
   C_{F}=\Big(1+\frac{\Upsilon}{\hbar fB}\Big)\log_{2}\Big(1+\frac{\Upsilon}{\hbar fB}\Big)-\Big(\frac{\Upsilon}{\hbar fB}\Big)\log_{2}\Big(\frac{\Upsilon}{\hbar fB}\Big)  \label{eq2}
\end{equation}
where $\hbar$ is the Planck’s constant and $f$ is the frequency of operation. 

Fock states cannot be legitimately described within the classical theory of electromagnetic radiation. To account for Fock states or any other non-classical states of light, the input states of light, their propagation and the output states of the detected signal need to be described in a way that transmission of classical information over a quantum channel is optimized. This is possible if the link is subject to Gaussian noise, as well as the input and output waveforms being Gaussian distributed. The quantum mechanical capacity over a Gaussian quantum channel carrying classical information without entanglement is upper bounded by the Holevo bound \cite{5}, 
\begin{align}
 C_{H} &=\Bigg(1+\frac{NB+\Upsilon \chi }{\hbar fB}\Bigg)\log_{2}\Bigg(1+\frac{NB+\Upsilon \chi}{\hbar fB}\Bigg)\nonumber\\ 
 &-\Bigg(\frac{NB+\Upsilon \chi }{\hbar fB}\Bigg)\log_{2}\Bigg(\frac{NB+\Upsilon \chi}{\hbar fB}\Bigg) \nonumber\\ 
 &- \Bigg(1+\frac{N}{\hbar f}\Bigg)\log_{2}\Bigg(1+\frac{N}{\hbar f}\Bigg)
  +\Bigg(\frac{N}{\hbar f}\Bigg)\log_{2}\Bigg(\frac{N}{\hbar f}\Bigg)
  \label{eq3}
\end{align}
where $N/\hbar f$ is the noise power spectral density and $\chi$ is the amplification/attenuation constant of the link, ($\chi < 1$; signal is attenuated, $\chi > 1$; signal is amplified). If the transmitter and the receiver pre-share entanglement as an additional resource, the capacity of a Gaussian quantum channel is given by the entanglement assisted Holevo bound \cite{6}, 
\begin{equation}
C_{E}=C_{F}(\zeta_{n})+C_{F}(\zeta_{n}+\eta_{\upsilon})-C_{F}(D_{+})-C_{F}(D_{-}) \label{eq4}
\end{equation}
in bits/sec where,
\begin{equation}
D_{\pm}=\frac{1}{2}(\sqrt{(2\zeta_{n}+\zeta_{\upsilon}+1)^2 - 4\chi\zeta_{n}(\zeta_{n}+1)} -1\pm\zeta_{\upsilon}), \nonumber
\end{equation}
$\zeta_{n}= \frac{N}{\hbar f}$ and $\zeta_{\Upsilon}= \frac{\upsilon\chi}{B \hbar f}$.

\section {Analytical Modelling}

The capacity bounds discussed so far consider that the case where the channel is linear with Gaussian distributed input, output and noise. The problem here is that the quantum limit to classical communication considers only the classical environment disregarding any quantum effects, while the quantum limit on quantum communication considers only the quantum environment disregarding classical channel effects. However, in a realistic communication network implementing fiber-optic or free-space optical quantum links carrying classical information, one will encounter both classical and quantum channel effects. Now, if the input and output waveforms follow arbitrary distributions, the resultant measurement will be subject to classical and quantum uncertainties. Let us consider that the arbitrarily distributed channel envelope is represented by the random variable $R$. The channel signal-to-noise ratio (SNR) can then be expressed as, $\gamma = R^2 \frac{\Upsilon}{N_{\beta\rho}(f)}$. For a continuous-input continuous-output linear arbitrarily distributed communication channel, $C$ can be derived in terms of, 
\begin{align}\label{eq5}
&C_{c-q-c} = \int_0^{\infty} B\log_2(1+\gamma)p_{\gamma}(\gamma)\mathrm{d}\gamma \nonumber\\
&\quad = \frac{2B\Upsilon}{N_{\beta\rho}(f)} \int_0^{\infty} r\log_2 \bigg(1 + \frac{r^2\Upsilon}{N_{\beta\rho}(f)}\bigg) p_R \bigg(\frac{r^2S}{N_{\beta\rho}(f)}\bigg) \mathrm{d}r                
\end{align}
where $N_{\beta\rho}(f)$ is the cross-power spectral density of Gaussian and quantum mechanical noise. Therefore, the crucial challenge to address here is to characterize $R$ and formulate $p_R (r)$.

\subsection{Approach}

We use the concept of coupled quantum harmonic oscillators to model the propagation environment in which a quantum link carrying either classical or quantum information operates. Using the analogy between the paraxial optical wave equation and the stationary Schrödinger equation \cite{7} and applying the Caldirola-Kanai Hamiltonian \cite{8} for solving the time-dependent Schrödinger equation; we calculate the propagation field strength and the corresponding average received signal energy. The amplitude of the paraxial wave can be assumed to be equivalent to the amplitude of a quantum harmonic oscillator. In that case, the intensity of the travelling paraxial wave will correspond to the probability distribution of the amplitude of the quantum harmonic oscillator. The phase component (Gouy phase \cite{9}) of the paraxial wave, in that case, will be equivalent to the time-dependency of the quantum oscillator.

Large-scale variations in the signal envelope can be determined by filtering signal energy with a moving average low pass filter operating over a time window within which large scale channel effects will remain unchanged. If the received signal energy is normalized by the square root of the moving average filter components, the resultant components will reflect the small-scale channel effects. Once the impact of the spatio-temporal variations in the received signal envelope over a realistic quantum link are captured, the distribution of R can be obtained. 

The noise spectral density, $N_{\beta\rho}(f)$ can be computed from Gaussian distributed classical noise and Poisson distributed quantum noise. With $p_R(r)$ and $N_{\beta\rho}(f)$ formulated, it then becomes possible to obtain $C_{c-q-c}$. If we consider the Gaussian and Poisson noises to be additive, we can use the summation of random variables to find the joint distribution of the hybrid noise. With the distribution known, it will be possible to calculate the mean, second central moment, variance, and spectral densities $N_{\beta\rho}(f)$ of the hybrid noise.To the best of our knowledge, this is the first-ever attempt to take the concept of quantum harmonic oscillators and apply it to characterize a quantum communication link carrying classical information. It connects the classical and the quantum worlds without assuming any propagation medium or its characteristics, input and output data formats, or transmitter and receiver structures.

\subsection{Methodology}

\subsubsection{Analogy between paraxial wave and stationary Schrödinger equations}

We start with the scalar Helmholtz equation for the electromagnetic vector $\mathbf{A}$ propagating through free space (vacuum) give by,
\begin{align}\label{eq5a}
\nabla^2 \mathbf{A} (\mathbf{r}, t) - \frac{1}{c^2} \frac{\partial^2 \mathbf{A} (\mathbf{r}, t)}{\partial t^2} = 0 
\end{align}
where $\mathbf{r} = \{x,y,z\}$ and $t$ are the position and the time coordinates of the vector $\mathbf{A}$, $c$ is the velocity of light in vacuum and $\nabla^2$ is the Laplacian operator. Now if we consider a monochromatic light beam with wave number $k$ is travelling only in the positive $z$ direction with only non-zero component $A_x$ in the $x$-direction, then we can write,
\begin{align}\label{eq5b}
A_x(\mathbf{r},t;k) = \psi(\mathbf{r})e^{\imath(kz - \omega t)}
\end{align}
where $\psi(\mathbf{r})$ is the electromagnetic (EM) field envelope and $\omega$ is the frequecny of operation. Putting (\ref{eq5b}) in (\ref{eq5a}) we can obtain, 
\begin{align}\label{eq5c}
\frac{\partial^2 \psi(\mathbf{r})}{\partial x^2} + \frac{\partial^2 \psi(\mathbf{r})}{\partial y^2} + 2\imath k \frac{\partial^2 \psi(\mathbf{r})}{\partial z} = 0.
\end{align}

We can solve (\ref{eq5c}) using the transverse EM mode $\psi_{lm}(\mathbf{r})$ that can be expressed as,
\begin{align}\label{eq5d}
\psi_{lm}(\mathbf{r}) =& \frac{w_0}{w(z)} \phi_l\bigg(\frac{\sqrt{2}x}{w(z)}\bigg) \phi_m\bigg(\frac{\sqrt{2}y}{w(z)}\bigg) e^{\frac{\imath k}{2R(z)}(x^2 + y^2)} \nonumber\\
&\times e^{-\imath(m+n+1)\phi(z)}
\end{align}
where $\phi_m(\beta) = \mathcal{H}_m(\beta)e^{-\beta^2/2}/\sqrt{2^m m!\sqrt{\pi}}$ is the harmonic oscillator wave function, $\mathcal{H}_m(\cdot)$ is the Hermite polynomial, $w_0$ is the minimal beam radius with Rayleigh range $b$, such that $w_0 = \sqrt{2b/k} = \sqrt{\lambda b/\pi}$ with $\lambda$ as the wavelength of operation and $w(z)$ is the beam radius at a distance $z$ from the source of the light. The corresponding longitudinal Gouy phase shift can be given by, $\phi(z) = \tan^{-1}(z/b)$ and the associated Gouy phase factor $e^{-\imath(m+n+1)\phi(z)}$ depends on the order $l$ and $m$ of the oscillator modes. Neglecting the transverse derivatives, we can find the electric filed vector using paraxial approximation as,
\begin{align}\label{eq5e}
\mathbf{E}(\mathbf{r},t; \mathbf{k}) = \Re\Big\{\Big\{\hat{\mathbf{x}}\omega \psi(\mathbf{r}) + \hat{\mathbf{z}}\imath c \frac{\partial \psi(\mathbf{r})}{\partial x}\Big\}e^{\imath(kz - \omega t)}\Big\}.
\end{align}

Next we apply coordinate transformation from $[x, y, z]$ to $[\xi, \eta, \tau]$ to move from the classical field to the quantum realm. Next we take the paraxial wave equation \cite{10},
\begin{align}\label{eq6}
2\imath k_0 \frac{\delta \mathbf{E}}{\delta z} = \nabla_{\perp}^2 \mathbf{E}+k^2 (x,y)\mathbf{E}  \end{align}
where $\nabla_{\perp}^2$ is the transverse Laplacian operator, $\lambda$ is the wavelength of operation, $k_0=2\pi n_0/\lambda$ is the wavenumber of the propagation mode with homogenous refractive index of $n_0$, $k^2(x,y)$ represents the inhomogeneity of the medium responsible for waveguiding the electromagnetic (optical) field $\mathbf{E}$, such that, $k^2 (x,y)= k_0^2-(k_x x^2+k_y y^2 ) + 2gxy$. The two displacement vectors $(x,y)$ physically represent the transverse and the longitudinal modes of the quantum field. The parameters $k_x$, $k_y$ and $g$ characterize the inhomogeneity of the medium through which the wave/particle travels. Next by putting $\psi(x,y,z) = \Psi(\xi,\eta;\tau)e^{\frac{\imath k}{2R(z)}(x^2+y^2)}/w(z)$ in (\ref{eq5c}), we can arrive at the Schrödinger equation for the quantum harmonic oscillator as,
\begin{align}\label{eq6a}
\Big[-\frac{\partial^2}{\partial \eta^2} - \frac{\partial^2 }{\partial \xi^2} + \eta^2 + \xi^2 - 2\imath \frac{\partial}{\partial \tau}\Big]\Psi(\xi,\eta;\tau) = 0.
\end{align}

Using the analogy between (\ref{eq6}) and Schrödinger equation \cite{7}, we can uncouple (\ref{eq6}) in two quantum harmonic oscillators. Using the concept of quantum harmonic oscillation, we can solve (\ref{eq6}) to formulate the propagation field as,
\begin{align}\label{eq7}
\mathbf{E}(x,y,z) &= e^{-\frac{\imath k_0 z}{2}} e^{\frac{\imath z}{k_0}(\omega_x + \omega_y)} e^{\frac{\imath z}{k_0}(\epsilon_x \omega_x + \epsilon_y \omega_y)} \nonumber\\
&\cdot\hat{\mathcal{W}}_{\theta}^{\dagger} \phi_{\epsilon_x}(x)\phi_{\epsilon_y}(y)    
\end{align}
where $z$ is the resultant displacement vector of $\mathbf{E}$, $\omega_x$ and $\omega_y$ are the phases associated with independent energy levels $\epsilon_x$ and $\epsilon_y$, $\phi_{\epsilon_x}(x)$ and $\phi_{\epsilon_y}(y)$ are the Hermite-Gauss functions or eigenfunctions of the harmonic oscillators and $\hat{\mathcal{W}}_{\theta}^{\dagger}$ is the conjugate transpose of the unitary operator, $\hat{\mathcal{W}}_{\theta}^{\dagger} = e^{i\theta(-i \frac{\delta x}{\delta y} + i \frac{\delta y}{\delta x})}$ and $\theta = \frac{1}{2}\tan^{-1}(\frac{2g}{k_x - k_y})$.

\subsubsection{Solving the time-dependent Schrödinger equation} 

The second part introduces the communication scenario where the transmitter and the receiver pre-share an unlimited amount of entanglement. We consider the scenario where the transmitter and the receiver both possess a cluster of particles. Particles belonging to each cluster are entangled with each other and we refer to it as intra-cluster entanglement. The entanglement shared between the transmitter and the receiver clusters is referred to as inter-cluster entanglement. Let $N_1$ and $N_2$ represent the photon numbers belonging to the transmitter and the receiver clusters, respectively. 

We start by rewriting (\ref{eq6}) in terms of Schrödinger equation with the Caldirola-Kanai Hamiltonian,
\begin{align}\label{eq7a}
\imath \hat{h} \frac{\partial \mathbf{E}'}{\partial z} = \hat{H}' \mathbf{E}'    
\end{align}
where $\hat{H}'= \hat{S}\hat{H}\hat{S}^{-1}$ and $\mathbf{E}' = \hat{S}\mathbf{E}$ with $\hat{S}$ and $\hat{S}^{-1}$ being the unitary and inverse unitary operators respectively, such that, $\hat{S}^{-1}\hat{S} = 1$. A weak solution for (\ref{eq7a}) will yield,
\begin{align}\label{eq7b}
\hat{S} = e^{\imath \gamma \frac{\partial}{\partial x}\frac{\partial}{\partial \eta}} e^{\imath \alpha \eta x}    
\end{align}
where $\alpha$, $\gamma$ are unknown constants chosen such that $\hat{H}'$ in (\ref{eq7b}) becomes diagonal. Proceeding in this way, we can solve for $\gamma$ and $\alpha$ to obtain, $\alpha = \sqrt{\omega_c/\omega} (\epsilon \pm \sqrt{\epsilon^2 + 1})$, $\gamma = \pm 1/2 \sqrt{\omega/\omega_c} 1/\sqrt{\epsilon^2 + 1}$, $\epsilon = \frac{\omega^2 - \omega_c^2 + \beta^2\omega}{2\beta\sqrt{\omega}\omega_c}$ where $\omega$ is the frequency of the classical information and $\omega_c$ is the frequency of the quantum harmonic oscillator, $\beta = \sqrt{4\pi/\omega V}$ with $V$ as the amount of quantization of the  classical information and $\epsilon$ is the independent energy level of the oscillator. Consequently, we can write,
\begin{align}\label{eq7c}
\hat{H}' = \frac{\omega}{2}\Big(\Lambda\eta^2 - \sigma \frac{\partial^2}{\partial \eta^2}\Big) + \frac{\omega_c}{2}\Big(\frac{1}{\sigma}x^2 - \kappa \frac{\partial^2}{\partial x^2}\Big)
\end{align}
where, $\Lambda = 1 + \alpha^2 \frac{\omega_c}{\omega} - 2\beta\alpha\frac{\sqrt{\omega_c}}{\omega} + \frac{\beta^2}{\omega}$, $\sigma = 1/(1 + \alpha^2 \frac{\omega}{\omega_c})$ and $\kappa = \sigma + \frac{2\beta\epsilon\gamma^2}{\sqrt{\omega}} - \frac{2\beta\gamma}{\sqrt{\omega_c}}(1 + \alpha\gamma)$. The solution for the non-stationary Schrödinger equation in the form, 
\begin{align}\label{eq7d}
\psi'(\xi, \eta, \tau) = \sum_{n_1, n_2} U_{n_1, n_2} e^{-i\epsilon_{n_1,n_2}\tau}.
\end{align}
Here $\psi'(\xi, \eta, \tau)$ can be obtained for,
\begin{align}\label{eq7e}
\hat{H}'\psi'_{n_1, n_2}(x, \eta) = \epsilon_{n_1,n_2}\psi'_{n_1, n_2}(x, \eta)
\end{align}
where $n_1$ and $n_2$ are photon numbers, $U_{n_1, n_2}$ are coefficients that depend on the initial conditions and $\epsilon_{n_1,n_2}$ are the energy levels of the Hamiltonian in (\ref{eq7a}) given by,
\begin{align}\label{eq7f}
\epsilon_{n_1,n_2} = \omega_c (n_2 + 1/2)\sqrt{G} + \omega (n_1 + 1/2)\sqrt{S}
\end{align}
where $G = \mp \frac{\beta\sqrt{\omega}}{2\sigma\omega_c\sqrt{\epsilon ^2 + 1}}$ and $S = 1 - \frac{\beta\omega_c}{\omega^{3/2}}(\epsilon \mp \sqrt{\epsilon^2 + 1}) + \beta^2/\omega$. The wave-function of the Hamiltonian can be calculated using the concept of separation as, $\psi'_{n_1, n_2}(x, \eta) = \psi'_{n_1}(x)\psi'_{n_2}(\eta)$ as,
\begin{align}\label{eq7g}
\psi'_{n_2}(x) &= C_{n_2} e^{-\frac{-R}{2}x^2} \mathcal{H}_{n_2}(x\sqrt{R}) \nonumber\\
\psi'_{n_1}(\eta) &= C_{n_1} e^{-\frac{-S}{2}\eta^2} \mathcal{H}_{n_1}(\eta S)
\end{align}
where $\mathcal{H}_{n_1}$ and $\mathcal{H}_{n_2}$ are Hermite polynomials, $R = \sqrt{{1}/{\sigma\kappa}}$, $S = \sqrt{{\Lambda}/{\sigma}}$ are the normalization coefficient, $C_{n_2} = R^{1/4}/\sqrt{2^{n_2}n_2!\sqrt{\pi}}$ and $C_{n_1} = S^{1/4}/\sqrt{2^{n_1}n_1!\sqrt{\pi}}$. Now in this scenario, the wave-function $\psi'_{n_1, n_2}(x, \eta)$ can be represented in form of a Fourier transform to obtain,
\begin{align}\label{eq7h}
\psi'_{n_2}(\eta) = \frac{1}{\sqrt{2\pi}} \int_{-\infty}^{\infty} a_{n_2}(\xi) e^{i\xi\sqrt{1/\sigma\kappa}\xi}\mathrm{d}\xi 
\end{align}
with $a_{n_2}(\xi) =  C_{n_2}(-\imath)^{n_2}e^{-\xi^2/2}$. As a result, we can write,
\begin{align}\label{eq7i}
\psi_{n_2,n_1}(x,\xi) =& C_{n_2}C_{n_1} (-\imath)^{n_2} \int_{-\infty}^{\infty} e^{-\xi^2/2} \mathcal{H}_{n_2}(\xi) e^{\imath x (\xi\sqrt{R} - \sigma\eta)} \nonumber\\
&\times e^{-S/2(\eta + \xi\gamma\sqrt{R})^2} \mathcal{H}_{n_1} (\sqrt{S}(\eta + \xi\gamma\sqrt{R})) \mathrm{d}\xi.
\end{align}
The integral in (\ref{eq7g}) can be calculated using table from \cite{11} to solve for $\hat{H}$ as, $i\frac{\delta \mathbf{E}}{\delta z} = \hat{H} \mathbf{E}$ where, 
\begin{align}\label{eq8}
\hat{H} =& \Bigg[\sum_{i = 1}^2 \sum_{j_i = 1}^{N_i}\frac{1}{2k_0}\big(\hat{p}^2_{x_{i,j_i}} + \hat{p}^2_{y_{i,j_i}}\big) \frac{1}{k_0}\bigg[\frac{k_0^2}{2} - \frac{1}{2}\bigg(k_{x_{i,j_i}}x^2_{i,j_i} \nonumber\\
&+ k_{y_{i,j_i}}y^2_{i,j_i} + g x_{i,j_i} y_{i,j_i}\bigg)\bigg]\Bigg] + \mu\bigg(\sum_{j_i1 = 1}^{N_1}\sum_{j_2 = 1}^{N_2} k_{x_{1,j_1}}\nonumber\\
&\times k_{y_{1,j_1}}k_{x_{2,j_2}}k_{x_{2,j_2}}\bigg) +\sum_{i = 1}^2 \sum_{j_i = m_i}^{N_i}\sigma_i x_{i,m_i} y_{i, m_i}
\end{align}
In (\ref{eq8}), $\hat{p}^2_{x_{i,j_i}}$ and $\hat{p}^2_{y_{i,j_i}}$ are the conjugate momentum operators, $k_{x_{i,j_i}}$ and $k_{y_{i,j_i}}$ are the displacement of the $j$th particle in the $i$th group, $\mu$ and $\sigma_i$ is the strength of entanglement between oscillators belonging to groups 1 and 2 and within group $i$ respectively. Using (\ref{eq8}), we extend our solution for $\mathbf{E}$ to the entanglement-assisted quantum communication case to obtain, 
\begin{align}\label{eq9}
&\mathbf{E}(x,y,z) = \sum_{i = 1}^2 \sum_{j_i = 1}^{N_i} e^{-\frac{ik_0 z}{2}} e^{\frac{iz}{k_0}(\epsilon_{x_{i,j_i}} \omega_{x_{i,j_i}} + \epsilon_{y_{i,j_i}} \omega_{y_{i,j_i}})} \nonumber\\
&\quad \cdot e^{\frac{iz}{k_0}(\omega_{x_{i,j_i}} + \omega_{y_{i,j_i}})} \hat{\mathcal{W}}_{\theta_{i,j_i}}^{\dagger} \phi_{\epsilon_{x_{i,j_i}}}(x_{i,j_i})\phi_{\epsilon_{y_{i,j_i}}}(y_{i,j_i})   \end{align}
where the unitary operator is $\hat{\mathcal{W}}_{\theta_{i,j_i}}^{\dagger} = e^{i\theta_{i,j_i}\big(-i \frac{\delta x_{i,j_i}}{\delta y_{i,j_i}} + i \frac{\delta y_{i,j_i}}{\delta x_{i,j_i}}\big)}$. 
\begin{figure*}[t]
\begin{center}
\includegraphics[width=1\linewidth]{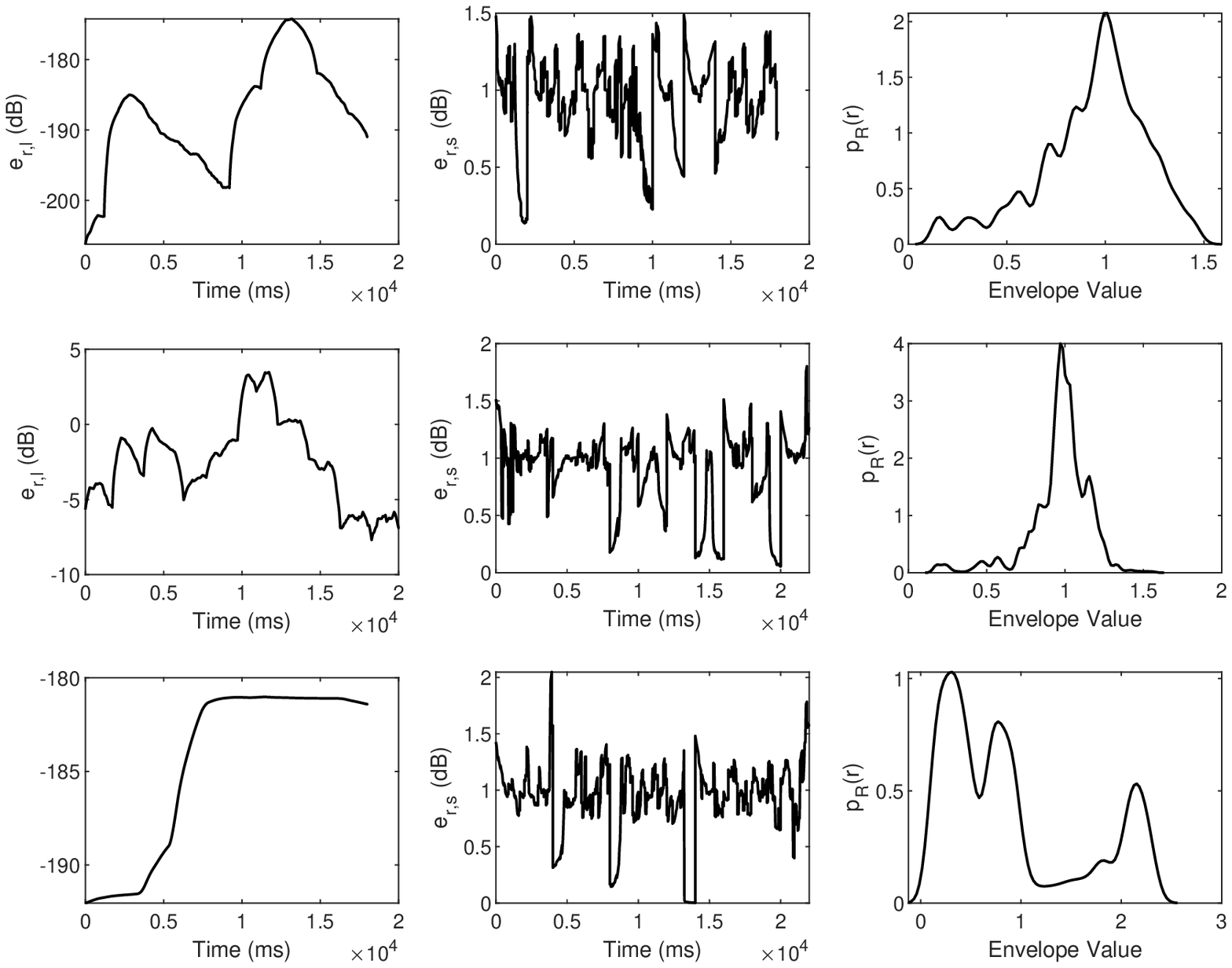}
\end{center}
\vspace*{-4mm}
\caption{Plots of large-scale variations in the signal envelope (left-most column), small-scale variations in the envelope (middle column) and distribution of the received signal envelope (right-most column) for  $\epsilon_{(x_{(i,j_i )})} = \epsilon_{(y_{ (i,j_i )})} = 2$ (top row),  $\epsilon_{(x_{(i,j_i )})} = \epsilon_{(y_{ (i,j_i )})} = 3$ (middle row) and $\epsilon_{(x_{(i,j_i )})} = \epsilon_{(y_{ (i,j_i )})} = 4$ (bottom row). For both cases, $k_{(x_{(i,j_i)})} = 1.2, k_{(x_{(i,j_i)})} = 1.5$ and $g = 0.25$.}
\label{fig1}
\vspace*{-2mm}
\end{figure*}
\begin{figure*}[t]
\begin{center}
\includegraphics[width=0.99\linewidth]{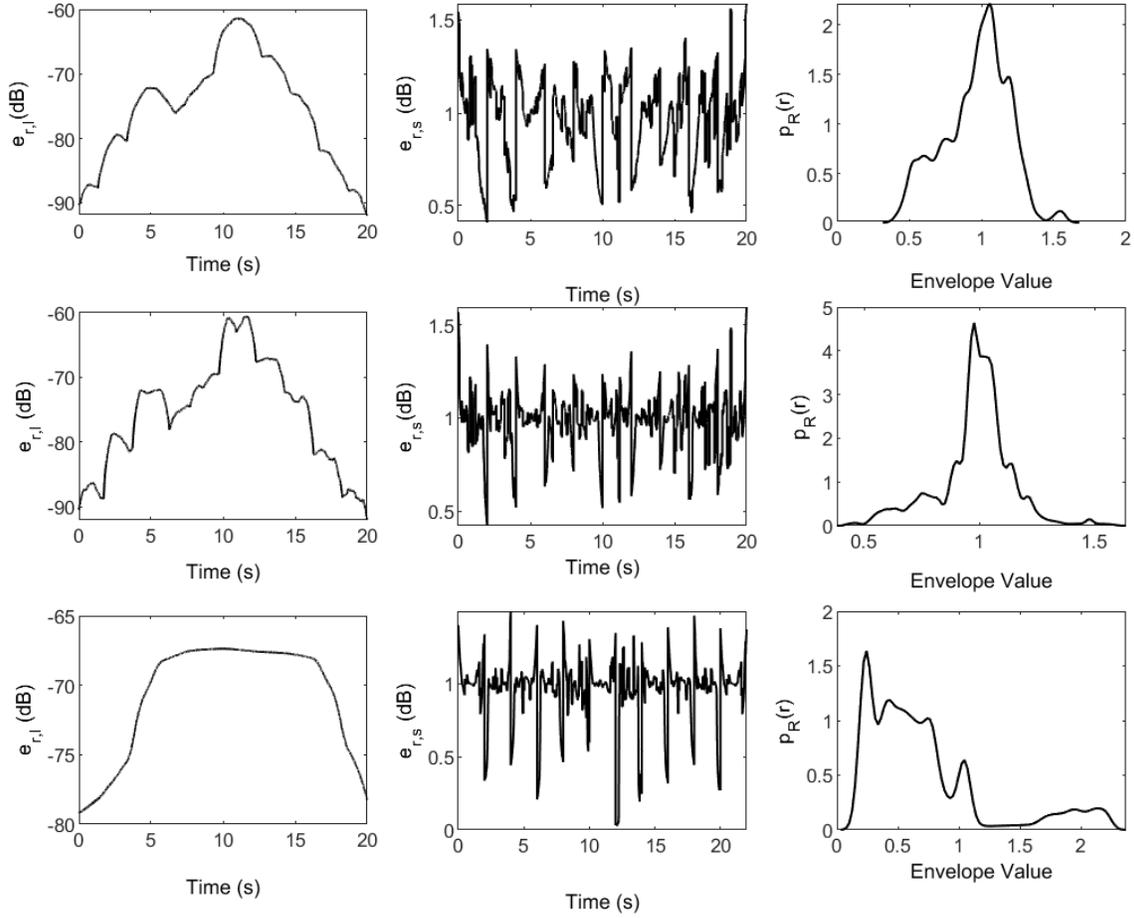}
\end{center}
\vspace*{-4mm}
\caption{Plots of large-scale variations in the signal envelope (left-most column), small-scale variations in the envelope (middle column) and distribution of the received signal envelope (right-most column) for  $\epsilon_{(x_{(i,j_i )})} = \epsilon_{(y_{ (i,j_i )})} = 2$ (top row),  $\epsilon_{(x_{(i,j_i )})} = \epsilon_{(y_{ (i,j_i )})} = 3$ (middle row) and $\epsilon_{(x_{(i,j_i )})} = \epsilon_{(y_{ (i,j_i )})} = 4$ (bottom row). For both cases, $k_{(x_{(i,j_i)})} = 3.5, k_{(x_{(i,j_i)})} = 5$ and $g = 0.5$.}
\label{fig2}
\vspace*{-2mm}
\end{figure*}

The third or final part introduces how the calculated electromagnetic field $\mathbf{E}$ can be used to calculate the distribution of the received signal envelope, $p_R (r)$. The solution $\mathbf{E}(x,y,z)$ from (\ref{eq9}) is plugged in to the following equation to calculate the average received signal energy, $\mathbf{e}_r = \frac{1}{2} \Re\{\mathbf{E}(x,y,z) \times \mathbf{E}^* (x,y,z)\}$, where $\Re$ denotes the real part of the formulation and $*$ denotes the complex conjugate. Next, large scale variations in the average received energy is determined by filtering signal energy $\mathbf{e}_r$ with a moving average low pass filter operating on a small time-window of 4$\lambda$s, over which large scale channel effects will remain approximately stationary. If the filtering process produces a vector of $\psi$, the square root of $\psi$ can be used to normalize $\mathbf{e}_r$ to get the small-scale variation in the signal envelope, $\mathbf{e}_{(r,s)}= \mathbf{e}_r/\sqrt{\psi}$. A sample set of small-scale variations observed in the signal envelope is presented on the second column of Fig.~\ref{fig1}.  Next a long averaging window of 150$\lambda$s is used for moving average filtering of $\psi$ to obtain a new set of filtering components $\psi '$ which is used to normalize $\psi$ to obtain the large-scale variations in the signal envelope, $\mathbf{e}_{(r,l)}= \psi/\psi '$.

\section{Numerical Results}

A sample set of large-scale variations observed in the signal envelope is presented on the first column of Fig.~\ref{fig1}. If $\mathbf{e}_{(r,s)}$ and $\mathbf{e}_{(r,l)}$ are modeled as random processes, the distribution of the received signal envelope, $p_R(r)$ can be obtained by multiplying the densities of $\mathbf{e}_{(r,s)}$ and $\mathbf{e}_{(r,l)}$. A sample set of $p_R(r)$  is presented on the third and last column of Fig.~\ref{fig1}. The results were generated in Matlab using high-performance computational facility. We start with the parameters, $k_x=1.2$, $k_y=1.5$ and $g=0.25$ for Fig.~\ref{fig1}. These set of values are commonly observed in graded-index (GRIN) media like multi-mode optical fibers, optical lenses etc. \cite{12}. We choose the energy levels, $\epsilon_x=\epsilon_y=\{2,3,4\}$ for the oscillator. They are the very basic energy levels or the least possible values that satisfy the Schrödinger equation. We assume also assume the wavelength of operation, $\lambda = 1300$ nm, and $N_i=4$ is the number of photons in each of the transmitter and receiver clusters that exhibit intra-cluster and inter-cluster entanglement. Plugging these parameters in (\ref{eq9}), propagation field $\mathbf{E}$ is computed, which in turn, is used to calculate the received signal energy, $\mathbf{e}_r$. Using moving average filtering technique mentioned before, large-scale, and small-scale variations, and distribution of the received signal envelope is extracted. Following the same procedure, another set of plots are generated in Fig.~\ref{fig2} for the large-scale variations, small-scale variations and distribution of the received signal envelope when the quantum oscillator energy levels are at $\epsilon_x = \epsilon_y = 2,3,4$. However, the equivalent parameters of the GRIN fibre are modified to $k_x = 3.5$, $k_y = 5$ and $g = 0.5$. 

It is evident from Figs.~\ref{fig1} and \ref{fig2}, that the communication scenario where classical or quantum information travels over a quantum field can be described in terms of the optical wave (which is also an EM wave) carrying information. This is due to the fact that photons are information carriers over quantum field while optical waves are physically made of millions of packets of photons. Therefore, it is possible to create an optical beam that starts out in the center and swings to either direction using the quantum harmonic oscillator language. By calculating the far-field distribution of received signal envelope over optical waves, it is possible to quantify the received signal strength over a quantum field carrying classical information. Since paraxial wave equation corresponds to the time-dependent Schrödinger equation for a two-dimensional harmonic oscillator, analogy between them can be used to understand the evolution of transmitted information (signal) over the quantum field in terms of the behavior of the harmonic oscillators. It is also possible to extend this concept and analysis to the case where a maximally entangled pair of photons are used to distribute information between the transmitter and the receiver.

Quantum entanglement results when the wave function of a system of quantum particles cannot be represented as a product of the wave functions of each particle. In order to characterize a quantum channel that can carry classical information where the information is teleported by a group of entangled quantum particles, we connected quantum entanglement with its space-time representation. We reformulated the two-dimensional quantum harmonic oscillator identifying the basic elements and the Hamiltonian independently for the classical system. In our representation, the ground states of the oscillators are maximally entangled and takes the form of one of the two Bell states $|\theta \rangle_{pm}$. It is worth-mentioning here that breaking the entanglement in vacuum may lead to the emission of quantum energy, a process referred to as the Hawking radiation \cite{13}.

\end{document}